\documentclass[12pt]{article}
\usepackage{amssymb}
\usepackage{color,graphicx}
\usepackage{amsmath}

\begin{document}

\title{Ambiguity in renormalization of X-junction between Luttinger liquid wires}
\author{D.N. Aristov$^{*}$ and  R.A. Niyazov\footnote{
Department of Physics, St.Petersburg State University, Ulianovskaya 1,
St.Petersburg 198504, Russia   and  NRC "Kurchatov Institute", Petersburg Nuclear 
Physics Institute, Gatchina 188300, Russia
}}
\date{}

\maketitle

\begin{abstract} 
\em We study four-lead junction of semi-infinite wires by using fermionic representation in the scattering state formalism. The model of spinless fermions with short-range Luttinger liquid type interaction is used.  We find  the renormalization group (RG) equations for the conductances of the system in the first order of fermionic interaction.  In contrast to the well-known cases of two-lead and three-lead junctions,  we show that the RG equations cannot be expressed solely in terms of conductances. The appearing ambiguity is resolved by choosing a certain sign defined at the level of S-matrix, but hidden in conductances.  The origin of this $Z_{2}$ symmetry is traced back to the particle-hole symmetry of our Hamiltonian. We demonstrate two distinct RG flows from any point in the space of conductances. The scaling exponents at the fixed points are not affected by these observations.
\end{abstract}


\section{Introduction}
 
One-dimensional (1D) systems of interacting fermions served as toy models for theorists since more than fifty years. 
\cite{GiamarchiBook}  The advances in technology and fabrication during last two decades propelled these models into the forefront of current research in physics at nanoscale. 
 Examples of such systems include carbon nanotubes,  \cite{Mebrahtu2013}
  1D edge states in quantum Hall regime \cite{Jezouin2013} and in topological insulators. \cite{Hasan2010, Qi2011}  One dimensional conductors and junctions between them should be important ingredients of any future electronic devices.
\cite{Halperin2012}   It is well known, that the transparency (conductance) of such junctions to the electric current is subject to renormalization due to interaction between fermions within the wire. For physical case of electronic repulsion this conductance tends to zero at low temperatures and applied voltages according to power law, with the exponent defined by the interaction strength. There are two approaches developed for the description of this conductance renormalization. 
 One approach uses the so-called bosonization technique, which treats the interaction between electrons in the bulk of the wire exactly, and considers the junction as a boundary. \cite{Kane1992} The system is described in terms of chiral fermionic densities (currents), and the boundary conditions are imposed onto these currents.  Fixed points of the boundary action are determined and perturbations around them are classified according to their scaling dimensions. \cite{Oshikawa2006,Hou2012} This way one judges about (in)stability of various fixed points, i.e. the limiting values of conductance obtained during renormalization procedure. 

Another approach to renormalization of conductance was first formulated in the limit of weak interaction in the bulk. In this approach one describes an arbitrary junction via its unitary S-matrix, defined in the absence of interaction. \cite{Yue1994} For $N$ semi-wires met at one junction this matrix belongs to $U(N)$ group.  The fermion wave function are given in a scattering state formalism, and corrections to S-matrix are subsequently calculated. \cite{Lal2002} These corrections are logarithmically divergent which eventually leads to the renormalization group (RG) equation for the  S-matrix and conductance.  This approach was essentially improved in \cite{Aristov2009,Aristov2013} by taking into account higher orders of interaction and subleading logarithms in perturbation theory for conductance. By summing up the appearing series one obtains the non-pertrubative RG equation for the  conductance, whose solutions nicely reproduce those scaling exponents which are known exactly from the bosonization approach. 

Up to now the main focus of theoretical analysis concerned the simpler cases $N=2$ and $N=3$, which describe an impurity in one wire and Y-junction with three leads, respectively.  The case of four-lead junction with $N=4$ received less attention, partly because of the difficulties in the description of $S$-matrix and ensuing analysis of RG equation even in the lowest order of perturbation theory.  We mention her recent studies  for a special model cases of $S$-matrix with $N=4$ while 
discussing regular networks of Luttinger liquids \cite{Kazymyrenko2005}, 
superconducting hybrid junctions \cite{Saha2013},  
and  tunneling between two helical edge states of the topological insulators \cite{Teo2009}. 

We notice that in all previous studies of two-lead and three-lead junctions the RG equations written in terms of $S$-matrix and in terms of conductance matrix (defined by squares of matrix elements $|S_{ij}|^{2}$) were equivalent. The same equivalence held in the considered simpler model cases of four-lead junctions as well.  
However,  there is no one-to-one correspondence for $N\ge4$  between the matrix of conductances and the $S$-matrix, even after removal of trivial phase factors. Mathematically, it is known as an ambiguity in restoring the unitary matrices from the unistochastic ones. \cite{Auberson1991,Dita1994}  In general, this ambiguity can  even be continuous and three-dimensional, while in our model below we only find a double discrete ambiguity which is a proven minimum for symmetric $S$-matrix in $U(4)$ group. \cite{Dita1994} 
   
The main goal of our paper is to demonstrate this ambiguity already in the first order RG equations. We show for a particular physically motivated class of $S$-matrices the description of the junction in terms of conductances is incomplete. We find two possible RG flows starting from any point in the conductances' space. The RG fixed points and the scaling exponents are not influenced by this ambiguity. 

The plan of the paper is as follows. We define our model in Sec.\ \ref{sec:model}, we explain the notion of reduced conductances in Sec.\ \ref{sec:conduc}, the RG equations are discussed in Sec.\ \ref{sec:RGeq}.
The ambiguity of the RG equations is revealed and discussed in Sec.\ \ref{sec:ambig}, we present our concluding remarks in Sec.\ \ref{sec:discus}.

\section{\label{sec:model} The model}

We consider a model of interacting spinless fermions describing two
quantum wires connected by a junction in the middle of the wires. Alternatively, we speak of 
four semi-wires connected at a single spot.  In the continuum limit,
linearizing the spectrum at the Fermi energy and including forward
scattering interaction of strength $g_{j}$ in wire $j$, we may write the Tomonaga-Luttiger liquid
Hamiltonian in the representation of incoming and outgoing waves in lead $j$
(fermion operators $\psi _{j,in}$, $\psi _{j,out}$) as

\begin{equation}
\begin{aligned}
\mathcal{H} &=\int_{-\infty }^{0}dx[H_{j}^{0}+H_{j}^{int}]\,,   \\
H^{0} &=v_{F}\Psi _{in}^{\dagger }i\nabla \Psi _{in}-v_{F}\Psi
_{out}^{\dagger }i\nabla \Psi _{out}\,, \\
H^{int} &=2\pi v_{F} \sum\limits_{j=1}^4 g_j\widehat{\rho }_{j}\widehat{\widetilde{\rho }}_{j}\Theta (x;-L,-l) \,.  
\end{aligned}
\label{eq:Ham}
\end{equation}

Here $\Psi _{in}=(\psi _{1,in},\psi _{2,in},\psi _{3,in},\psi _{4,in})$ denotes a vector
operator of incoming fermions and the corresponding vector of outgoing
fermions is expressed through the $S$-matrix as $\Psi _{out}(x)=S\cdot \Psi
_{in}(-x)$ . In the chiral representation we are using positions on the
negative (positive) semi-axis corresponding to incoming (outgoing) waves. We
consider quantum wires of finite length $L$, contacted by reservoirs. The
transition from wire to reservoir is assumed to be adiabatic (i.e. produces
no additional potential scattering). The junction is assumed to have
microscopic extension $l$ of the order of the Fermi wave length. Interaction effects inside the
junction are neglected. This is expressed by the window
function $\Theta (x;-L,-l)=1$, if $-L<x<-l$, and zero otherwise. The regions $x<-L
$ are thus regarded as reservoirs or leads labeled $j=1,2,3,4$. We put the Fermi
velocity $v_{F}=1$ from now on.  
The interaction term of the Hamiltonian is expressed in
terms of density operators $\widehat{\rho }_{j,in}=\Psi ^{+}\rho _{j}\Psi =%
\widehat{\rho }_{j}$, and $\widehat{\rho }_{j,out}=\Psi ^{+}\widetilde{\rho }%
_{j}\Psi =\widehat{\widetilde{\rho }}_{j}$, where $\widetilde{\rho }%
_{j}=S^{+}\cdot \rho _{j}\cdot S$ and the density matrices are given by $%
(\rho _{j})_{\alpha \beta }=\delta _{\alpha \beta }\delta _{\alpha j}$ and $(%
\widetilde{\rho }_{j})_{\alpha \beta }=S_{\alpha j}^{+}S_{j\beta }$. 
The  $S$-matrix describes the scattering at the junction and belongs to $U(4)$ group.  

 \begin{figure}[h] \begin{center}
 \begin{minipage}[h]{0.47\linewidth}
 \includegraphics[width=1\textwidth]{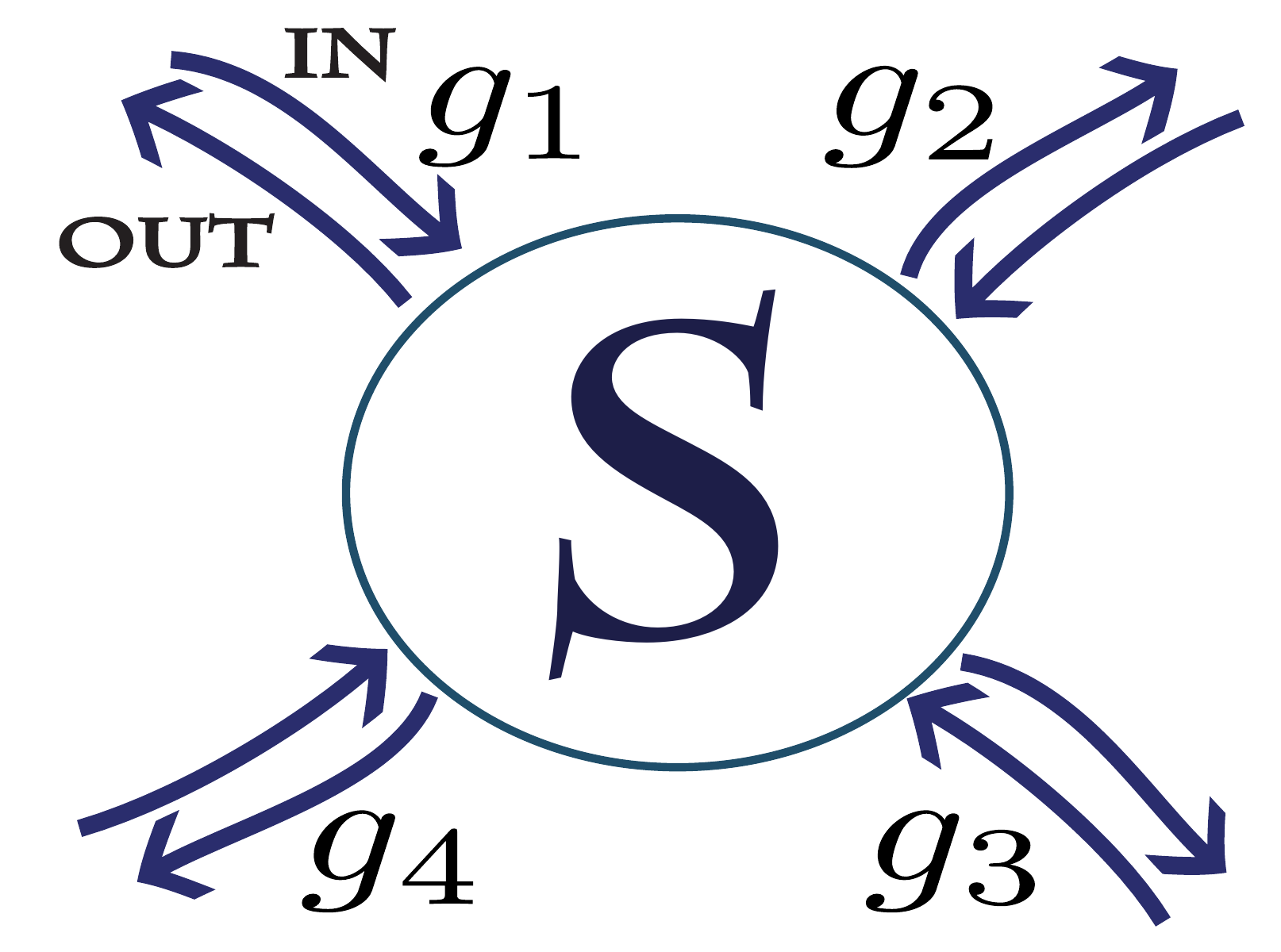}
 \caption{A four-lead junction of quantum wires, corresponding to the Hamiltonian \eqref{eq:Ham}. 
 \label{fig:s-matrix}}  
 \end{minipage}  \hfill    \begin{minipage}[h]{0.47\linewidth}
 \includegraphics[width=1\linewidth]{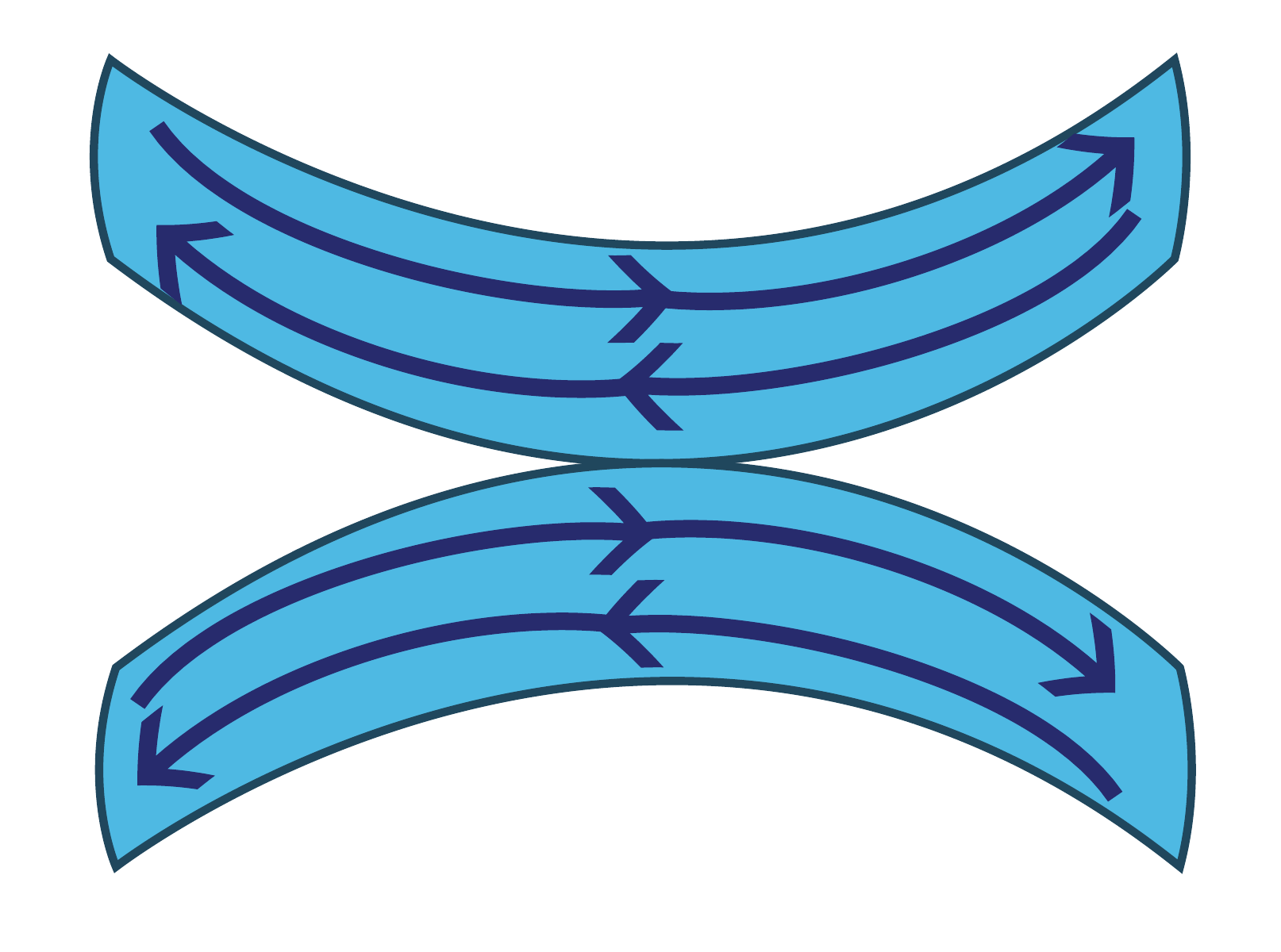}
 \caption{Point contact of two wires is schematically shown, illustrating discrete symmetries in our choice of $S$-matrix, Eq.\ \eqref{Smatrix}. \label{fig:2wires}}
 \end{minipage} \end{center}  \end{figure}

\section{\label{sec:conduc} Reduced conductances}

In the linear response regime our system is characterized by the matrix of conductances defined by $I_i=C_{ij} V_j$, with  the current $I_i$ flowing in wire $i$ and the  voltage $V_j$ applied to the wire $j$.  The current conservation, $\sum I_i=0$, and the absence of response to the equal change in voltages result in  
the Kirchhof's rules, $\sum _{i}C_{ij} = \sum _{j}C_{ij} = 0$. 
It suggests that we can choose more convenient linear combinations of  $I_i$, $V_{j}$ reducing the number of independent components in $C_{ij}$.  In the d.c.\ limit  we have from the Kubo formula $C_{ij}=\frac12 (\delta_{ij}-Y_{ij}) $, with $Y_{ij} = |S_{ij}|^{2}$ ; one can also write $Y_{ij} = \mbox{Tr}(\widetilde \rho_{i} \rho_{j})$.  \cite{Aristov2011a}  

The appropriate representation for the reduced conductance matrix may be constructed by using 
  generators of $U(4)$ Cartan subalgebra, which are three traceless diagonal matrices and one unit matrix. We define 
\begin{equation}
\begin{aligned}
\mu_1 &=\mbox{diag}(1,-1,0,0), \quad
\mu_2 =\mbox{diag}(0,0,1,-1),\\
\mu_3 &= 1/\sqrt{2} \ \mbox{diag}(1,1,-1,-1),\quad
\mu_4 =1/\sqrt{2} \ \mbox{diag}(1,1,1,1).
\end{aligned}
\end{equation}
with the property  $\mbox{Tr}(\mu _{j}\mu _{k})=2\delta _{jk}$,  $j=1,\ldots, 4$. The densities
are  expressed  as $\rho _{j}=1/\sqrt{2}\sum_{k }R_{jk }{%
\mu }_{k }$, where the 4$\times$4 matrix $\mathbf{R}$ is given by
\begin{equation}
\mathbf{R}= \begin{pmatrix}
 \frac{1}{\sqrt{2}} & 0 & \frac{1}{2} & \frac{1}{2} \\
 -\frac{1}{\sqrt{2}} & 0 & \frac{1}{2} & \frac{1}{2} \\
 0 & \frac{1}{\sqrt{2}} & -\frac{1}{2} & \frac{1}{2} \\
 0 & -\frac{1}{\sqrt{2}} & -\frac{1}{2} & \frac{1}{2} 
\end{pmatrix}
\end{equation}%
and has the properties $\mathbf{R}^{-1}=\mathbf{R}^{T}$, $det\,\mathbf{R}=1$. 
The outgoing amplitudes are expressed in  a similar form with 
$\mu_{j}$ replaced by $\widetilde{\mu }_{j}=S^{+}\cdot \mu _{j}\cdot S$. 
It also means \cite{Aristov2011} that we work now with the  combinations of currents and voltages of the form
$I_{i} =1/\sqrt{2} \sum_{k} R_{ik} I_{k}^{new}$,  $V_{i} = 1/\sqrt{2} \sum_{k} R_{ik} V_{k}^{new}$,  or 
\begin{equation}
\begin{aligned}
I_{1}^{new} &= (I_{1} - I_{2})/2  \,,\quad  V_{1}^{new} = (V_{1} - V_{2})/2  \,,  \\ 
I_{2}^{new} &= (I_{3} - I_{4})/2  \,,\quad  V_{2}^{new} = (V_{3} - V_{4})/2 \,,  \\ 
I_{3}^{new} &=( I_{1} + I_{2} - I_{3} - I_{4} )/\sqrt{8} \,,  \\
 V_{3}^{new} &= (V_{1} + V_{2}  -V_{3} - V_{4}) /\sqrt{8}  \,,  \\ 
 I_{4}^{new} &=\sum\nolimits_{j} I_{j}/\sqrt{8} \,,  \quad 
 V_{4}^{new}  = \sum\nolimits_{j} V_{j} /\sqrt{8} \,, 
\end{aligned}
\end{equation}
We could use more physical combinations $V_{1} - V_{2}$ etc. without additional factors 1/2, $1/\sqrt{8}$,  see \cite{Aristov2011a}, however, it is irrelevant for our purposes below. 

The  reduced conductance matrix in such basis is determined by $\mathbf{G}=\mathbf{R}\,\mathbf{C}\, \mathbf{R}^T$ and has a structure
\begin{equation}
\mathbf{G} =
\begin{pmatrix}
3\times3 & 0 \\
 0 & 0
\end{pmatrix}
\end{equation}
In the presence of interactions, the structure of the last expression is unchanged, but the elements vary. 
The main effect in the d.c. limit can be described by the renormalization of the $S$-matrix,  \cite{Aristov2011a} 
which translates to the renormalized quantity,   
\begin{equation} \label{eq:yr}
Y^R_{ij}=\tfrac{1}{2} \mbox{Tr}(\widetilde{{\mu }}_{i}^{\;r}{\mu} _{j})
\end{equation}
where the superscript $R$ shows that we work in the basis $\mu_{j}$ instead of $\rho_{j}$, and the superscript $r$ denotes that the quantity is fully renormalized by interactions. From now on we assume that all quantities are renormalized and drop this latter superscript. 

\section{\label{sec:RGeq} Renormalization group equations}

The renormalization of the conductances by the interaction is determined by
first calculating the correction terms in each order of perturbation theory in $g_{j}$.
We are in particular interested in the scale-dependent contributions
proportional to $\Lambda =\ln (L/l)$, where $L$ and $l$ are two lengths,
characterizing the interaction region in the wires, Eq.\ \eqref{eq:Ham}. 
The above expression for $\Lambda$ corresponds to vanishing  temperature, $T\ll v_{F}/L$, and for higher temperatures we should replace $\Lambda =\ln (l^{-1}/\max[L^{-1},T/v_{F}])$.
In lowest order of perturbation the scale dependent contribution to the
conductances is given by \cite{Aristov2012a}
\begin{equation}
G_{jk}=G_{jk}\big\rvert_{\mathbf{g}=0}
+\tfrac{1}{2}\sum\limits_{l,m} \mbox{Tr}\left[\widehat{W}_{jk}\widehat{%
W}_{lm} \right] g_{ml}\Lambda \,,
\label{RG1order}
\end{equation}%
where  $G_{jk}\rvert_{\mathbf{g}=0}=\delta _{jk}-Y_{jk}$, the $\widehat{W}_{jk}= [\rho_{j}, \widetilde\rho_{k}]$ are a set of
sixteen $4\times 4$ matrices (products of $\widehat{W}$'s are matrix products), 
$g_{ml}=g_{m}\delta _{ml}$ is the matrix of interaction constants and the
trace operation $\mbox{Tr}$ is defined with respect to the 4$\times$4 matrix space
of $\widehat{W}$'s. 

We  multiply $G_{ij}$ by $\mathbf{R}^{T}$ from the left and by $\mathbf{R}$ from the
right to get the components of $\mathbf{Y}^{R}$ in the form
\begin{equation}
Y_{jk}^{R}=  Y_{jk}^{R} \big\rvert_{\mathbf{g}=0}-\frac{1}{2}\sum\limits_{l,m} 
\mbox{Tr}\left[\widehat{W}_{jk}^{R}\widehat{W}_{lm}^{R}\right]
g_{ml}^{R}\Lambda \,.
\end{equation}
Differentiating these results with respect to $\Lambda $ (and then putting $%
\Lambda =0$) we find the RG equations in the first order in the interaction
\begin{equation}\label{eq:rgeq}
\frac{d}{d\Lambda }Y_{jk}^{R}=-\frac{1}{2}\sum \limits_{l,m}
\mbox{Tr}\left[\widehat{W}_{jk}^{R}\widehat{W}_{lm}^{R}\right] g_{ml}^{R} \,.
\end{equation}%
The number of non-zero matrices $\widehat{W}_{jk}^{R}=\{\mathbf{R}^{T}\cdot \widehat{\mathbf{W}}
\cdot \mathbf{R}\}_{jk}$ is reduced to nine in most general case, since  $\widehat{W}_{j4}^{R}= \widehat{W}_{4j}^{R}=0$ ; we have  $g_{ml}^{R}=\{\mathbf{R}^{T}\cdot \mathbf{g} \cdot \mathbf{R}\}_{ml}$. The  matrices 
$\widehat{W}_{jk}^{R}$ are best evaluated with the aid of computer algebra.

\section{\label{sec:ambig} RG flow ambiguity phenomenon}

We continue our analysis by appropriate choices of interaction $g_{ij}$ and S-matrix parametrization.
Let interaction constants first be equal in both wires
$  \mathbf{g}  = \mbox{diag}(g,g,g,g) $,
and we parametrize the $S$-matrix by three angles in the following way
\begin{equation}
S=\left(
\begin{array}{cccc}
r_1 & t_1 & f_2 & f_2 \\
t_1 & r_1 & f_2 & f_2 \\
f_1 & f_1 & r_2 & t_2 \\
f_1 & f_1 & t_2 & r_2
\end{array}
\right), 
\label{Smatrix}
\end{equation}
where
\begin{equation}
\begin{aligned}
r_1 & = \frac12 (e^{-i \alpha_1}+ \cos \beta  ), \quad
t_1 = \frac12 (-e^{-i \alpha_1}+ \cos \beta  ) , \\ 
r_2 & = \frac12 (e^{-i \alpha_2}+ \cos \beta  ), \quad
t_2 = \frac12 (-e^{-i \alpha_2}+ \cos \beta  ) , \\  
f_1&=  f_2=\frac i 2 \sin  \beta  .
\end{aligned}
\end{equation}
This parametrization \eqref{Smatrix} describes the transmission and reflection in each wire  ($\alpha_1$, $\alpha_2$) and the hopping between the wires ($\beta$). The above Eq.\ \eqref{RG1order} is invariant upon ``rephasing'', i.e. the multiplication of $S$ from both sides by unitary matrices of the form $\mbox{diag}(e^{i\gamma_{1}},e^{i\gamma_{2}},e^{i\gamma_{3}},e^{i\gamma_{4}})$. Without loss of generality we may assume $\beta \in [0,\pi/2]$, $\alpha_{1,2} \in [-\pi, \pi]$.

The computation of the reduced conductances matrix~\eqref{eq:yr} yields 
\begin{equation}
\mathbf{Y}^R 
= \begin{pmatrix}
\cos \beta \cos  \alpha_1 & 0 & 0 \\
 0 & \cos  \beta \cos  \alpha_2 & 0 \\
 0 & 0 & \cos ^2  \beta \\ 
 \end{pmatrix}
 \equiv \begin{pmatrix}
a_1 & 0 & 0 \\
 0 & a_2 & 0 \\
 0 & 0 &b \\ 
 \end{pmatrix}
 \label{def:YR}
\end{equation}
The matrix form of the RG equation \eqref{eq:rgeq} is now written as a set of coupled RG equations for components of $\mathbf{Y}^R $  in terms of the initial variables  
\begin{equation}
\begin{aligned}
\frac{d Y^R_{11}}{d\Lambda}&=
 g \left(\tfrac{1}{4}\sin ^2\beta \left(\cos \alpha _1(2 \cos \beta+\cos \alpha _2 )-3\right) 
 \right. \\ &\left.+1-\cos ^2\alpha _1 \cos ^2\beta-\tfrac{1}{4}\sin ^2\beta\sin \alpha _1 \sin \alpha _2 \right), \\
\frac{d Y^R_{22}}{d\Lambda}&= 
\frac{d Y^R_{11}}{d\Lambda} \Big| _{\alpha_1 \leftrightarrow \alpha_2} \,, \\
\frac{d Y^R_{33}}{d\Lambda}&= g \sin ^2\beta \cos \beta \left(\tfrac{1}{2}(\cos \alpha _1+\cos \alpha _2)+ \cos \beta \right)  \,.
\end{aligned}
\label{RGamb}
\end{equation}
Our natural desire is to express these equations entirely in terms of the conductances, as it was successfully done in our previous studies for two and three wires connected by the junction \cite{Aristov2009,Aristov2010,Aristov2013}.  

We have three independent components of the reduced conductances matrix, denoted as $a_1$, $a_2$, $b$ in Eq.\ \eqref{def:YR}. The attempt to write the right-hand side of the RG equations \eqref{RGamb} in terms of $a_1$, $a_2$, $b$  faces the ambiguity problem in the term $\propto \sin  \alpha_1 \sin   \alpha_2$. The sign of this term depends on angles range: if $\alpha_1$ and  $\alpha_2$ both belong to the range either $(0,\pi)$ or $(-\pi,0)$  then $\sin  \alpha_1 \sin   \alpha_2$ is positive, but if  $\alpha_1$ and $\alpha_2$ belong to different segments $(0,\pi)$  and  $\alpha_2$ to $(-\pi, 0)$ then the discussed term is negative.  Notice, that the values of conductances $a_{1}$ and $a_{2}$ are not affected by the change of sign $\alpha_{1}\to - \alpha_{1}$ and $\alpha_{2}\to - \alpha_{2}$, respectively.   This change of sign corresponds to complex conjugation of some elements of $S$-matrix \eqref{Smatrix}, namely, $r_{i}\to r_{i}^{*}$ and  $t_{i}\to t_{i}^{*}$, which cannot be compensated by ``rephasing'' operations. 

One may ask what is the internal discrete symmetry, manifesting itself at the level of RG equations? To answer this, we consider two decoupled ($\beta=0$) Luttinger liquids with impurities. Standard calculations \cite{Aristov2009} show that (at least in the lowest Born approximation) the phase $\alpha_{j}$ is equal to $U_{bs}/v_{F}$, with $U_{bs}$ the backscattering amplitude off the impurity in $j$th wire. 
The sign of $U_{bs}$ is unimportant, as only the square $|U_{bs}|^{2}$ defines the conductance \cite{Kane1992}. When allowing hopping between the wires, we start to feel the difference between two cases: i) the scattering potentials in both wires are of the same sign, i.e. both bumps or both pits, or ii) scattering potentials in the wires are of different sign, i.e. a pit in one wire and a bump in another. 
Another explanation of the sign question in Eq.\ \eqref{RGamb} concerns the particle-hole symmetry of the Hamiltonian \eqref{eq:Ham}. At the level of decoupled wires the sign of $U_{bs}$ (and $\alpha_{j}$) is changed when passing to hole description, $\Psi_{in} \to \Psi_{in}^{\dagger}$, etc. One may then regard the above sign question as arising from the possibility to perform the particle-hole transformation in one wire. 

We see that the RG equations \emph{cannot} be defined in terms of conductances only. 
Generally, we have two different RG flows for conductances, and the choice between them should be done on the base the initial phases of $S$-matrix,  $\alpha_1$, $\alpha_2$. 
Further analysis of RG equations shows that the ambiguity doesn't influence the position of fixed points (FPs). We have four FPs parametrized by $a_1=\pm1$, $a_2=\pm1$,  $b=1$, which reads as $\alpha_{1,2} = 0,\pi$, $\beta=0$ in terms of angles. These FPs correspond to simple cases of two separated wires with absolute transmission or reflection in each of them. The fifth FP is defined by $a_1= a_2= b=0$ and discussed below. 

To clarify the character of different FPs we generalize our consideration and allow for different interaction constants in  two wires :
\begin{equation}
 \mathbf{g}  = \mbox{diag}(g_1,g_1,g_2,g_2).
\end{equation}
The representation of the RG equations purely in terms of conductances is cumbersome and we rewrite them in terms of the angles \cite{Aristov2011a} 
\begin{equation}
\begin{aligned}
\frac{d\alpha_{1}}{d\Lambda} = & -\frac{1}{4\cos\beta}   \left(g_1 (1+3  \cos^{2}  \beta ) \sin
   \alpha _1 -g_2  \sin ^2\beta \sin \alpha _2  \right) 
   \\
\frac{d\alpha_{2}}{d\Lambda} =   &\frac{d\alpha_{1}}{d\Lambda} 
\Big| _{\alpha_1 \leftrightarrow \alpha_2 , g_1 \leftrightarrow g_2} \,,
 \\
 \frac{d\beta}{d\Lambda} =
   & -\frac{1}{4} \sin \beta
   \left(g_1 \cos \alpha _1  + g_2 \cos \alpha
   _2  +   (g_1+g_2 ) \cos \beta \right) \\
\end{aligned}
\label{RGeqAngles}
\end{equation}
The same ambiguity of RG equations in terms of conductance is seen again in \eqref{RGeqAngles}. One can  change, e.g.  $\alpha_{2} \to -\alpha_{2}$, without changing the conductance but altering the RG flows. 

As before, we observe four universal FPs,  $\alpha_{1,2} = 0,\pi$, $\beta=0$ (i.e.\ $a_{1,2}=\pm1$,  $b=1$). Only one of these four is a stable FP and it is defined by the sign of the interaction in individual wires. According to usual expectations \cite{Kane1992,Yue1994} we have for the stable FP: $a_{j} = \mbox{sign } g_{j}$.  In addition we find a fifth non-universal  FP, which  is never stable, and whose position attains a compact form in terms of conductance: 
\begin{equation}
a_1  =-a_{2}= \frac{g_2-g_1}{g_1 + g_2},  \quad b=a_1^2 \,.  
\label{FP5}
\end{equation}
This FP is in physical domain at $|a_{1}| < 1$, which happens for $g_{1}g_{2} > 0$. 

\begin{figure}[h!] 
\begin{minipage}[h]{0.47\linewidth} 
\center{\includegraphics[width=1\linewidth]{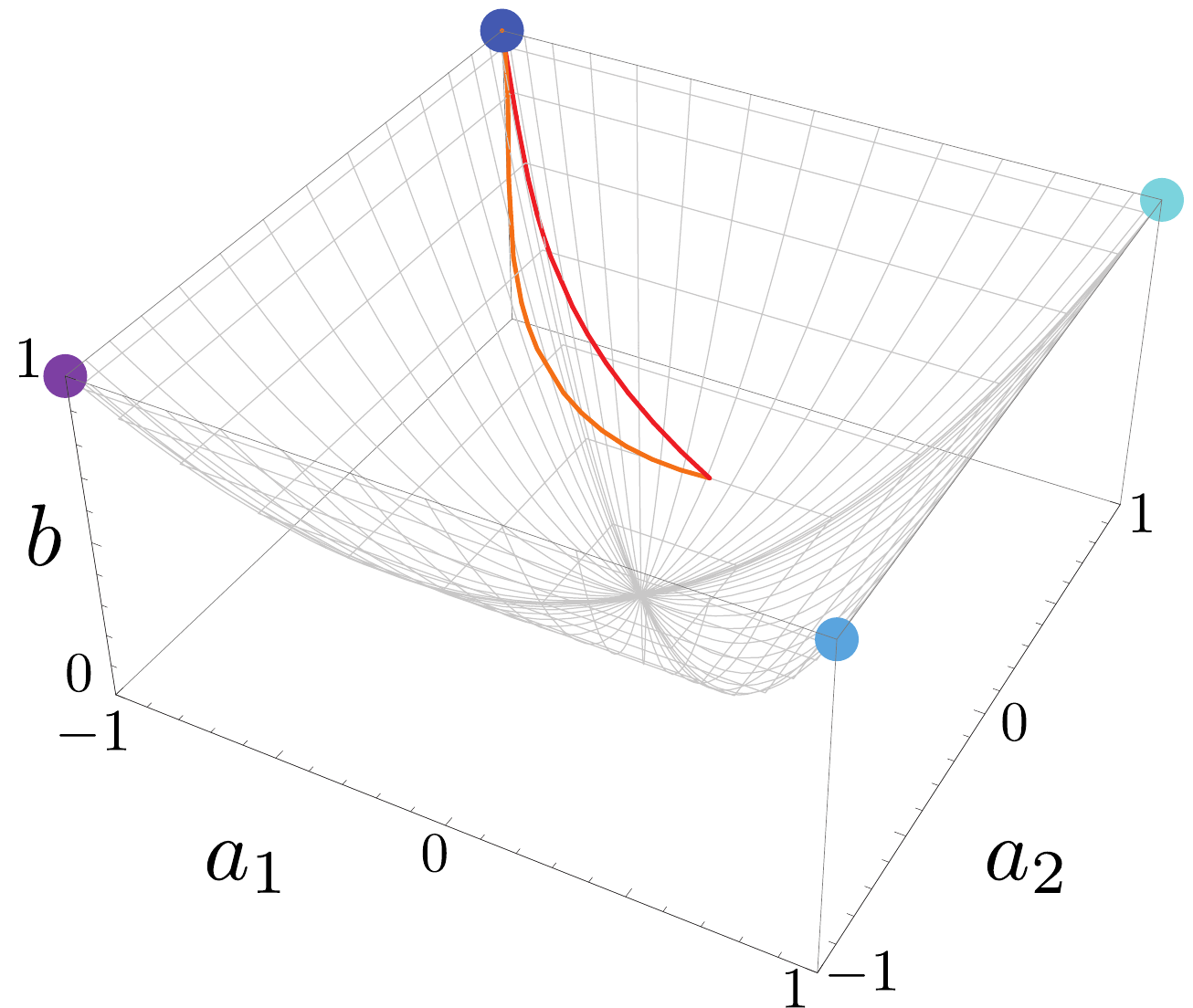}} a) $g_1=-0.4,\ g_2=0.2$ \\
\end{minipage}
\hfill
\begin{minipage}[h]{0.47\linewidth}
\center{\includegraphics[width=1\linewidth]{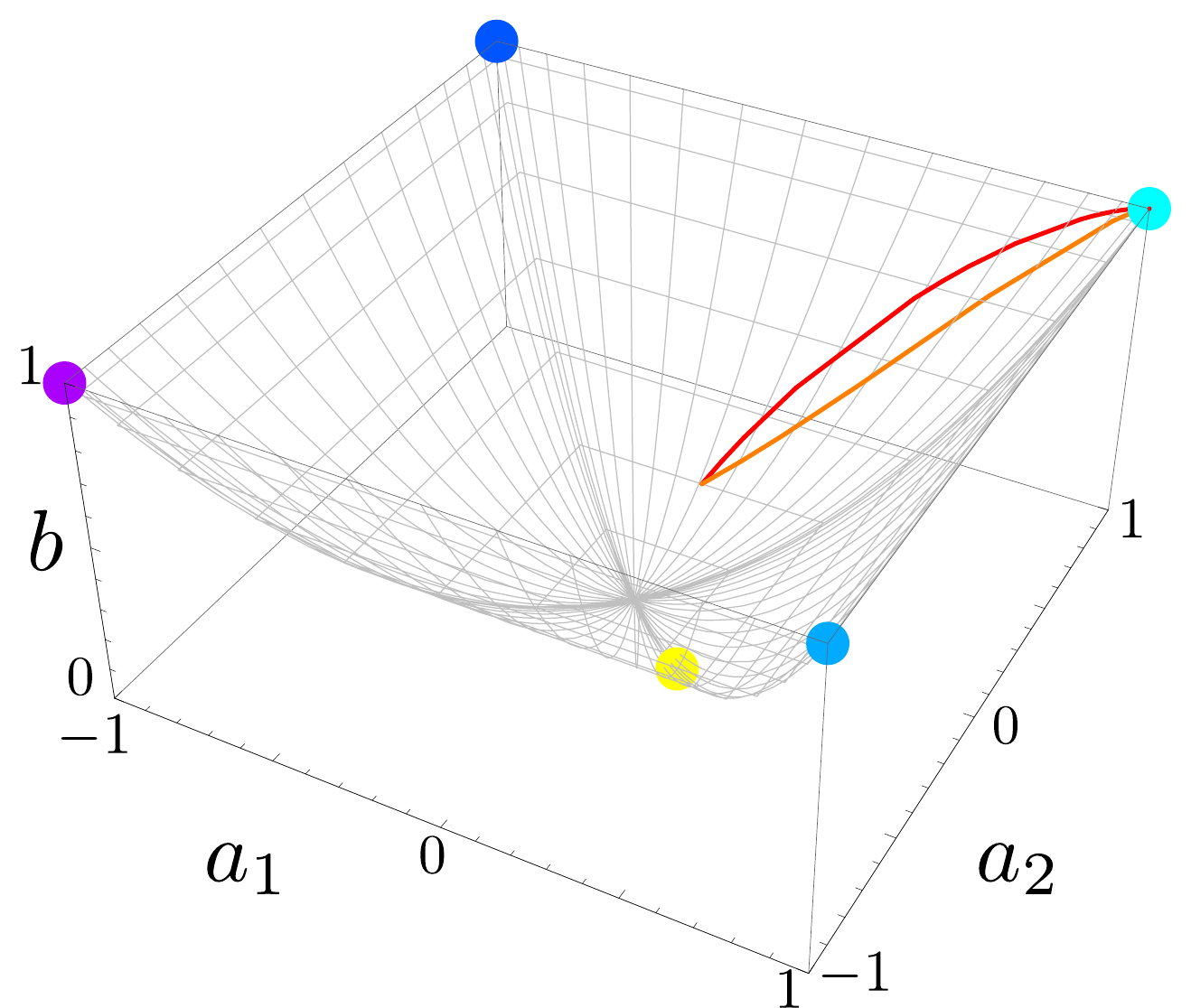}} b) $g_1=0.37,\ g_2=0.12$  \\
\end{minipage}
\vfill
\begin{minipage}[h]{0.47\linewidth}
\center{\includegraphics[width=1\linewidth]{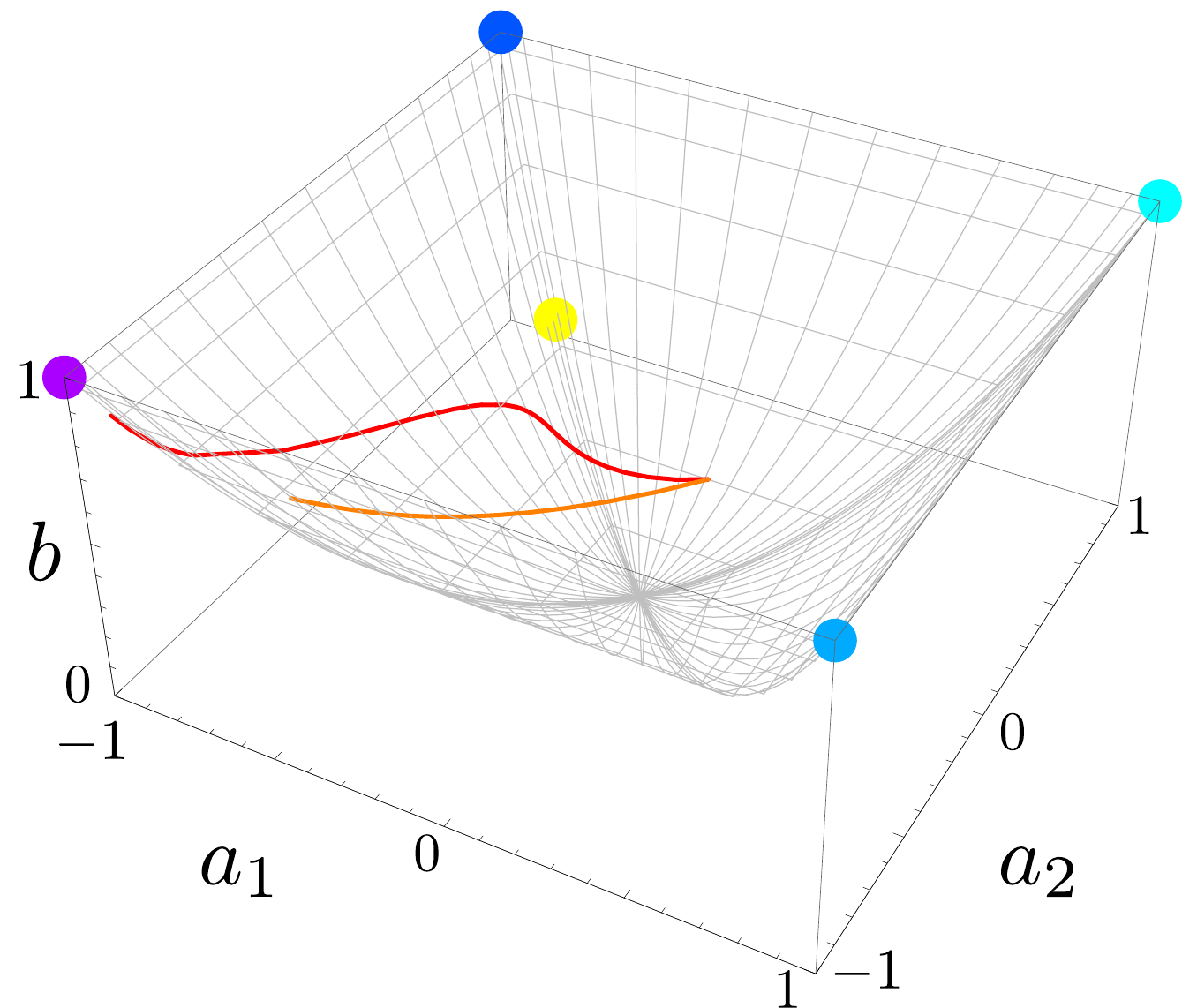}}  c) $g_1=-0.4,\ g_2=-0.1$  \\
\end{minipage}
\hfill
\begin{minipage}[h]{0.47\linewidth}
\center{\includegraphics[width=1\linewidth]{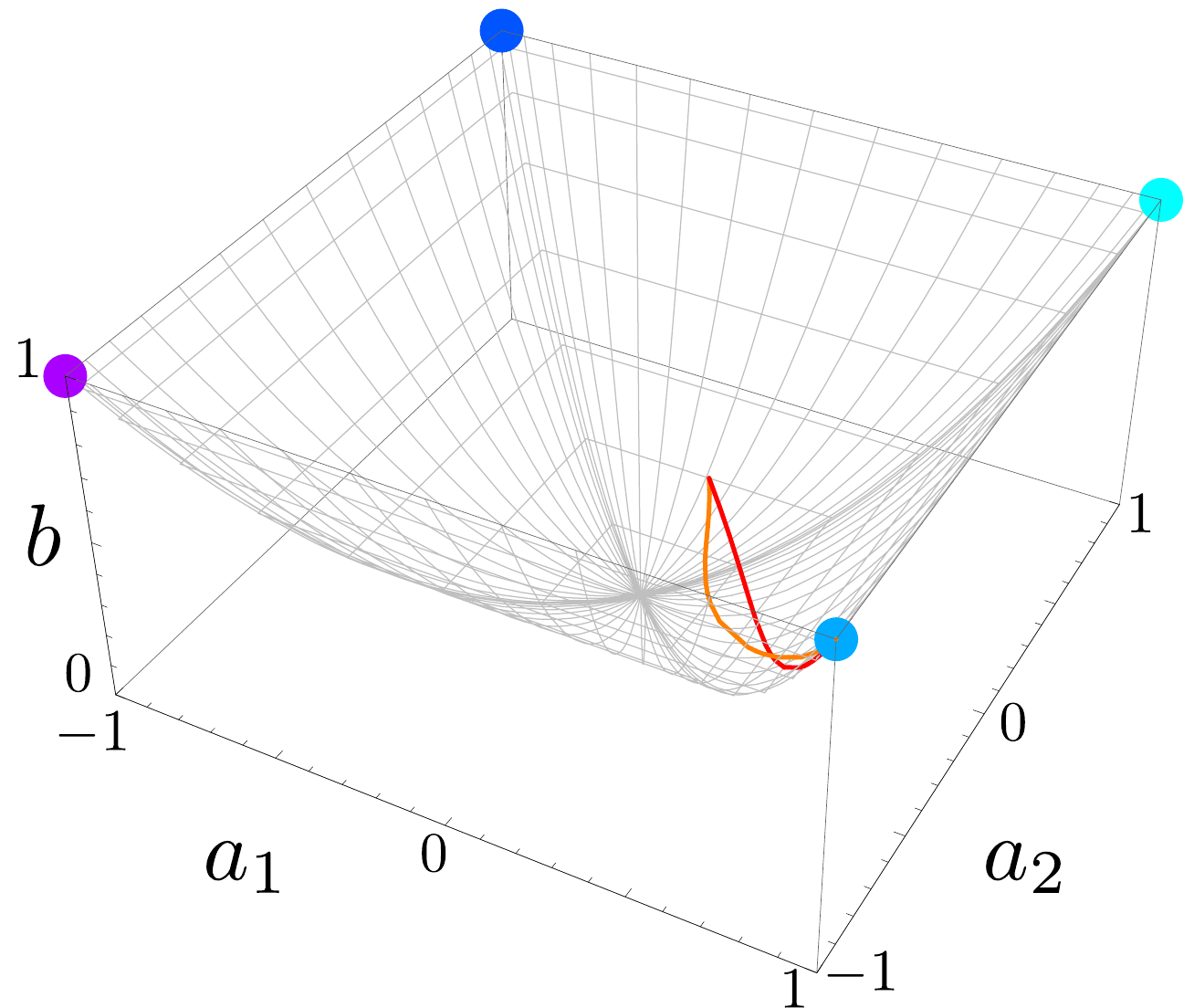}}  d) $g_1=0.21,\ g_2=-0.28$  \\
\end{minipage}
\caption{\label{fig:FPandRG}
Fixed points and RG trajectories in conductances space. Two possible trajectories lead to the attractive FP, determined by the signs of interaction in wires. It is seen that the RG flows of the conductances are non-monotonous functions of the scaling variable. The fifth non-universal FP appears when $g_{1}g_{2} > 0$, its position defined by Eq.\ \eqref{FP5}.  
}
\end{figure}

We illustrate our findings in Fig.~\ref{fig:FPandRG}, where  we show the body of conductances, i.e. the set of allowed conductances values $(a_{1},a_{2},b)$ and  possible RG trajectories, starting from the same parametric point $(a_{1},a_{2},b)$ for different values of  $g_1,g_2$.  We see that the stable FP depends on the quarter in the plane of $(g_1,g_2)$, similar to the situation with Y-junction \cite{Aristov2012a}. We also demonstrate the existence of two possible RG flows, leading to \emph{the same} stable FP (darker one is for plus sign in ambiguity term, lighter - for minus).  One can verify that the scaling exponents near the FPs are not affected by the discussed sign ambiguity. Two possible RG flows result only in a different prefactors in the scaling dependence of the conductances. For instance, for the repulsive interaction in both wires, $g_{1}, g_{2} >0$, if the RG flow starts in the vicinity of FP $|\alpha_{1}| , |\alpha_{2}|, \beta \ll 1$ then we have  
\begin{equation}
1-a_{1,2} \sim  \beta^{2} e^{-(g_{1}+g_{2})\Lambda}  +
\alpha_{1,2}^{2} e^{-2g_{1,2}\Lambda},  \quad 
1-b \sim  2\beta^{2} e^{-(g_{1}+g_{2})\Lambda}   
\label{scaling}
\end{equation}
Starting far away from such vicinity, two possible RG trajectories will end at the different points $\alpha_{j},\beta$ in \eqref{scaling} but the scaling law is the same. 
 
\section{\label{sec:discus} Discussion}
In this paper we study four-lead junction of spinless Luttinger liquid wires by fermionic representation in scattering state formalism.  The interaction in wires leads to the renormalization of the conductances of the system, expressed via the absolute values of the $S$-matrix elements. The RG equation for the conductances can usually be formulated entirely in terms of conductances, which was explicitly checked for general two-lead and three-lead junctions.  In this paper we demonstrate that in case of four-lead junctions the RG equations possess the sign ambiguity unresolved in terms of conductances. From a mathematical viewpoint, this ambiguity is an intrinsic property of $U(4)$ group, and is most easily seen in the sign choice, which defines the left- and right-isoclinic subgroups of $SO(4) \subset U(4)$. From a physical viewpoint, the ambiguity may be traced back to the particle-hole symmetry of our Hamiltonian. This results in two possible non-monotonous dependences of the renormalized conductances as functions of the scaling variable. 

We note that should we use the bosonization approach in our analysis, we would not observe the ambiguity in question. This is because the bosonization starts with the FPs of RG equations and analyzes the scaling dimensions of perturbations around it. We show above that the scaling dimensions (exponents) do not depend on the choice of RG flow, and it is only prefactors before the scaling exponentials which are determined by the particular RG trajectory, leading to the close vicinity of FP from the distant points in parameter space. 
 
We believe that the discussed ambiguity may be an important issue in further theoretical investigation and experimental manipulation of X-junctions between quantum wires.

{\bf Acknowledgments.}  We are grateful to P. W\"olfle for useful discussions.  This research 
 was supported by the Russian Scientific Foundation grant (project 14-22-00281).

\end{document}